\documentclass[twoside,leqno,twocolumn]{article}  
\usepackage{ltexpprt}     
\usepackage{graphicx}
\usepackage{times}
\usepackage{mathfont}
\def\log{\mathop{{\rm log}}}

\def\max{\mathop{{\rm max}}}

\makeatletter
\def\subsection{\@startsection{subsection}{2}{0pt}{-12pt}{3pt}{\normalsize\bf}
} 
\makeatother

\DeclareSymbolFont{lasy}{U}{lasy}{m}{n}
\let\Box\undefined
\DeclareMathSymbol\Box{0}{lasy}{"32}
\makeatletter
\let\old@endproof\@endproof
\def\@endproof{\hfill$\Box$\old@endproof}
\makeatother

\mathcode`O="724F

\begin{document}

\title{\Large Internet Packet Filter Management and Rectangle Geometry}

\author{David Eppstein\thanks{Dept. Inf. \& Comp. Sci., Univ. of
California, Irvine, CA 92697-3425.  Email: {\tt
eppstein@ics.uci.edu}. Work performed in part while visiting AT\&T.} \\
{\small Univ. of California, Irvine}
\and
S. Muthukrishnan\thanks{AT\&T Labs, Shannon Laboratory, 180 Park
Ave., Florham Park, NJ 07932. Email: {\tt muthu@research.att.com}.} \\
{\small AT\&T Labs Research}}

\date{}
\maketitle

\pagestyle{myheadings}
\markboth{}{}

\begin{abstract}
We consider rule sets for internet packet routing and filtering,
where each rule consists of a range of source addresses, a range of
destination addresses, a priority, and an action.  A given packet
should be handled by the action from the maximum priority rule that
matches its source and destination.  We describe new data structures for
quickly finding the rule matching an incoming packet, in near-linear
space, and a new algorithm for determining whether a rule set contains
any conflicts, in time $O(n^{3/2})$.
\end{abstract} 

\bigskip

\section{Introduction}

The working of the current Internet and its posited evolution
depend on efficient packet filtering mechanisms: databases of rules,
maintained at various parts of the network, which use patterns to filter
out sets of IP packets and specify actions to be performed on those
sets. Typical filter patterns are based on packet header information such
as the source or destination IP addresses. The actions to be performed
depend on  where the packet filtering is performed in the network. For
example, at backbone routers, packet filters specify which interface or
link to use when forwarding packets. In firewalls, packet
filters specify whether to allow a connection. More generally, packet
filters specify Quality-of-Service actions such as restricting certain
classes of traffic to no more than a threshold bandwidth. This packet
filtering mechanism --- maintaining a database of  filters with
associated actions and applying them to IP packets as appropriate ---
underlies most crucial aspects  of the Internet: correct routing,
providing security,  guaranteeing service level agreements between
different subnets, billing based on  traffic patterns, etc.

Implementing the packet filtering mechanism in the Internet
involves sophisticated packet filter management tasks. 
In particular, we need {\em packet classification}, that is, given 
an IP packet with a specific header values, we need to determine which 
filter applies to that packet. We also need {\em filter conflict 
detection}, that is, we need to determine whether two or more filters that
apply to a packet specify conflicting actions. Conflicts 
are resolved by adding additional filters, so the filter database
remains consistent. These are the fundamental 
packet filter management tasks governing the IP network performance.

In this paper, we present efficient algorithms for solving 
both of these packet filter management problems. Our approach is to
solve the underlying abstract problem which, in each case, 
is naturally formulated as a geometric data structural problem.
We focus on simple techniques suitable for highly efficient
implementations, especially in our packet classification algorithms,
because in the future we hope to explore  implementations of them in
practical applications. However our work provides theoretical asymptotic
improvements as well.

The same abstract geometric data structural problems
derived from these packet filtering applications
arise independently in other important applications areas as well,
and our results improve
the best known results for those applications.
In what follows, we describe the packet filter management 
problems (Section \ref{sec:problems}) 
and our results (Section \ref{sec:results}), 
and provide an overview of our techniques (Section \ref{sec:tech})
before providing all the details (Sections \ref{sec:classify} to
\ref{sec:conflict}). We will briefly describe the other application
areas where our results are relevant in Section \ref{sec:tech}.

\subsection{Packet Filter Management Problems}
\label{sec:problems}

A packet filter $i$ in IP networks is a collection of
$d$-dimensional  {\em ranges} $[l_i^1,r_i^1] \times \cdots \times
[l_i^d,r_i^d]$,  an {\em action} $A_i$ and a {\em priority} $p_i$.  
The precise nature of action is not relevant here except 
that we can determine if two actions $A_i$ and $A_j$ 
are in {\em conflict} (for example, if $A_i$ is to allow
the packet through the firewall and $A_j$ is to disallow 
it, there is a conflict of action). Any IP packet $P$ can be viewed as
a $d$-dimensional vector of  values $[P_1,\ldots,P_d]$
summarizing the header information of the packet.
A filter $i$ {\em applies} to packet 
$P$  if $P_j \in [l_i^j,r_i^j]$ for each $j\in [1,d]$. 

\begin{description}
\item{{\bf Packet Classification Problem.}}
A database $F$ of filters is available for preprocessing. 
Each online query is a packet $P$, and the goal is to {\em classify}
it, that is, to 
determine the filter of highest priority that applies to $P$. 
A related problem is to list all filters that apply to $P$. 

\vfil\eject

\item{{\bf Filter Conflict Detection Problem.}}
Given a database $F$ of filters, determine if there exists
any packet $P$ such that of the filters of the highest 
priority that apply to $P$, any two of them specify 
actions that conflict. Reated problems are to list all
regions wherein conflicting $P$'s lie, or to list all 
conflicting pairs of filters. 
\end{description}

Some remarks follow. 
Existing IP routers use destination based routing, 
that is, use filters with $d=1$ specifying ranges of destination
IP addresses. As the Internet evolves from being the best effort
network as it is now to provide differentiated services, 
two or more IP header fields may be specified by a filter.
Some proposals are underway to specify many fields such as
source IP address, destination IP address, source port, destination
port etc., while others are underway which seem to preclude 
using more than just the source and destination IP addresses, that is,
$d=2$ (in IPsec for example, the source or destination port
numbers may not be revealed.)  In the rest of this paper, 
we will assume $d=2$  and the fields that are specified 
are source and destination IP addresses since that seems 
likely to be most prevalent and of immediate interest.

Filters typically specify IP address ranges 
as an IP address $a_1\cdots a_{32}$ 
and a mask of certain number $l$ of bits, that is, 
the range is $a_1\cdots a_{l}00\cdots 0$ to 
$a_1\cdots a_{l}11\cdots 1$. So these are 
not arbitrary ranges. Instead they are {\em hierarchical},
that is, if two ranges intersect, one is completely
contained in the other. All our results will in fact
work for arbitrary ranges in each dimension although 
some of our algorithms can be made simpler for implementation
purposes if the ranges are hierarchical.  

In both problems we will let $n$ denote the number of filters 
in $F$. The value of $n$ varies depending on where filtering 
is done: backbone routers may have hundreds of thousands of filters,
firewalls may only have a few hundreds, etc.  
All numbers are integers in the range $[0,U-1]$ --- for
IP addresses, this is currently $[0,2^{32}-1]$, but may go up to $2^{64}$
or higher in IPv6. 

\subsection{Our Results}
\label{sec:results}

Our main results are as follows. 
\begin{itemize}
\item
{\em Packet Classification Problem.} We present 
an algorithm for this problem with different tradeoffs 
for data structure space vs filtering time. In particular,
we obtain very fast classification times with near-linear space: with
$O(n^{1+o(1)})$ space, classification takes $O(\log\log n)$ time, or with
$O(n^{1+\epsilon})$ space, classification takes $O(1)$ 
time.\footnote{For clarity, we have  stated the results for $U=n^{O(1)}$;
bounds for general
$U$ appear later in the paper. Throughout this section, we assume 
$U=n^{O(1)}$ for making comparisons with existing results.}

\item
{\em Filter Conflict Detection Problem.}
We present an $O(n)$ space, $O(n^{3/2})$ time algorithm for 
this problem. Straightforward $O(n^2\log n)$ time algorithms
were the only known previous result.
\end{itemize}
 
The packet classification problem has been extensively 
studied with over a dozen papers in the premier networking 
conferences (INFOCOM and SIGCOMM)
in the past few years (e.g., see references in~\cite{FelMut-INFOCOM-00}).
Classification time is of paramount importance
(for example, for backbone routers,
filtering IP packets has to be done at the speed at which it 
forwards the packets, a blistering speed!). 
However, at such high speeds, memory is very expensive
and the consensus in the networking community is that
classification must be very fast, but 
that data structural space must be limited to the extent possible.
The applied works in INFOCOM and SIGCOMM
use near-linear space, but take time $\Omega(\log n)$ 
to classify each packet which they attempt to further speed up
using large memory cache line etc. However, the golden standard 
has been the bound of $\Theta(\log\log n)$ that can be achieved  
for the $d=1$. With the 
exception of~\cite{FelMut-INFOCOM-00}, known algorithms for 
$d=2$ fail to meet this bound. Our algorithmic result here meets this
bound, but uses only $O(n^{1+o(1)})$ space improving upon the 
$\Theta(n^{1+\epsilon})$ space needed 
by~\cite{FelMut-INFOCOM-00,FerMutdeB-STOC-99} 
which is the previously best known result.  Furthermore, our
result is easily implementable; hence, it 
additionally holds promise as a practical packet classification
solution. 

The filter conflict detection problem has received attention
only recently~\cite{AdiSurPar-INFOCOM-00}. That work was primarily
motivated  by detecting security  holes in firewalls. Filter databases
in firewalls get modified by systems administrators manually or
automatically (for example, when a host from inside a firewall requests
a TCP connection with a host outside,  a filter may be added to the
firewall to enable the  target host to  open a TCP connection through the
firewall).  Conflicts arise quite naturally, and the task of the
administrator is to resolve them appropriately. The work
in~\cite{AdiSurPar-INFOCOM-00} was motivated by this scenario. However, 
conflict detection helps in auditing filter
databases~\cite{DecDitPar-ToN-00} in general for ambiguities in routing,
unfulfilled service guarantees etc., that is, in general where packet
filter mechanism is employed.  It is straightforward to solve this
problem in $O(n^2\log n)$ time. Our main contribution here is in breaking
the quadratic barrier and designing an $O(n^{3/2})$ time algorithm. 

\subsection{Our Techniques and Other Applications\\ of Our Results}
\label{sec:tech}

Both the packet classification and the filter 
conflict detection problem can be thought of as 
geometric problems in which each rule is a $2$-dimensional axis-parallel
rectangle.\footnote{In $d$ dimensions,
they will be $d$-dimensional hyperrectangles.}
The packet classification problem can be viewed as locating
a point (the query IP packet header values) in a partition of space
formed by overlaying these rectangles.
The filter conflict detection problem is that of detecting 
certain overlapping regions among rectangles of highest 
priority that overlap a region. Our approach is to solve 
the underlying geometric data structural problems in the bounds 
quoted above. This has other immediate applications, for example
to the problems in~\cite{FerMutdeB-STOC-99}, giving the following 
new results: 
(1) faster multi-method lookup in object oriented languages
and the first known efficient algorithm for auditing multi-method
libraries,
(2) improved matching algorithms for rectangular matching,
wherein, for the first time, matching time is independent
of the dictionary size while the space used is subquadratic
in dictionary size, and (3) the first known optimal algorithm for
approximately matching a pattern string with edit distance at most $1$ in 
a text -- the matching time is linear in the text size and 
preprocessed space is sublinear in the dictionary size. 
These three problems have extensive literature, and 
all these results are of independent interest. Readers are 
referred to~\cite{FerMutdeB-STOC-99} for details. 

Our approach to solving the two packet classification relies on a
standard plane-sweep approach to turn the static two-dimensional
rectangle query problem into a dynamic one-dimensional problem,
in which we maintain a dynamic set of intervals and must again
query the maximum priority set element containing a query point.
This one-dimensional problem must be solved {\em persistently},
so we can query previous versions of the data structure after
the plane sweep has occurred.  We solve this persistent one-dimensional
problem using a data structure combining ideas from B-trees and segment
trees.

Our approach to the filter conflict detection problem uses a technique
related to an algorithm by Overmars and Yap~\cite{OveYap-SJC-91} for {\em
Klee's measure problem} (determining the volume of a union
of rectangular blocks): we use a kD-tree~\cite{Ben-CACM-75} to divide
the plane into rectangular cells, not containing any rectangle vertex, so
that the rectangles intersecting any cell form {\em stripes} (i.e.,
rectangles that are unbounded in one dimension).  The conflict detection
problem can thus be reduced to determining a lower envelope of line
segments, which can also be interpreted data structurally as an offline
priority queue problem or graph theoretically as a minimum spanning tree
verification problem.  We solve this subproblem efficiently using a
linear-time union-find data structure.

\section{Fast Packet Classification Queries}
\label{sec:classify}

As described above, packet classification can be viewed as an orthogonal
range querying problem, in which we wish to find the maximum priority
rectangle containing any query point.  We now describe data structures
for solving this problem efficiently.

\subsection{Persistent Interval Queries}

First, we consider a dynamic one-dimensional query problem:
what is the maximum priority interval containing a query point
among a dynamically changing set of intervals, having integer endpoints
in the range $[0,U-1]$.  We assume without loss of generality that $U$ is
a power of two. Our data structure will be {\em partially persistent}:
an update must be performed on the most recent version of the structure,
but a query can refer to any prior version.

Our data structure will be parametrized by a value $k$,
and will consist of {\em blocks} of $O(2^k)$ memory words,
each corresponding to information about an interval of values
within the range $[0,U-1]$.
An update may create new blocks but will not change existing blocks.
If a block corresponds to query values in the interval
$[x,y]$, then by {\em subinterval $i$} we refer to the interval
$[x+i(y-x)2^{-k},x+(i+1)(y-x)2^{-k}-1]$.
A persistent version of the data structure will be represented by a
pointer to a block forming the top level of the data structure.

Each block contains the following information:
\begin{itemize}
\item A table ${\rm opt}[i]$ of pointers to the maximum-priority
interval in the dynamic set that contains subinterval $i$.
\item A table ${\rm pq}[i]$ of pointers to priority queue data
structures for the intervals containing subinterval $i$.
\item A table ${\rm subblock}[i]$ of pointers to blocks
representing the subset of dynamic intervals having endpoints in
subinterval $i$.  If a subinterval contains no endpoints, this pointer
is null.
\end{itemize}

The priority queues are not used for queries, and so do not need to be
maintained persistently.  We will later see how to eliminate them
altogether for problems derived from hierarchical rectangle sets.

\begin{lemma}
The data structure described above can find the maximum priority
interval containing a query point in time $O((\log U)/k)$.
\end{lemma}

\begin{proof}
We answer a query simply by repeatedly following the pointer ${\rm
subblock}[i]$ for the subinterval~$i$ that contains the query point.
For each block found via this chain of pointers, we look up the value
${\rm opt}[i]$ and compare the priorities of the intervals found in this
way.

Each successive block in the chain corresponds to an interval of size
smaller by a $2^{-k}$ factor than the previous block, so the total
number of blocks considered is $\log_{2^k} U=(\log U)/k$.
For any interval ${\cal I}$ containing the query point there is a
maximal block such that ${\cal I}$ contains the subinterval
containing the query in that block; then by the assumption of maximality
${\cal I}$ must have an endpoint in the block and is a candidate for
${\rm opt}[i]$.  Therefore, the true maximum-priority interval
containing the query is one of the ones found by the query, and the
query algorithm is correct.
\end{proof}

\begin{lemma}
The data structure described above can be updated in time
$O((2^k\log n)(\log U)/k)$
and space $O(2^k(\log U)/k)$.
\end{lemma}

\begin{proof}
To insert or delete an interval,
we create a new copy of each block containing one of the endpoint
intervals.  By the same argument used to bound query time, there
are at most $2(\log U)/k$ such blocks.
For each copied block, we update the priority queues corresponding to
subintervals containing the updated interval, copy pointers to these
priority queues into the ${\rm pq}[i]$ pointers of the new block, and use
these priority queues to set each value of
${\rm opt}[i]$.  We then copy each pointer ${\rm subblock}[i]$
from the previous version of the block, except for the one or two
subintervals containing the updated interval's endpoints, which are
changed to point to the new blocks for those subintervals.
Each update causes the creation of at most $2(\log U)/k$ new blocks,
using space $O(2^k(\log U)/k)$.  Each update also changes
$O(2^k(\log U)/k)$ priority queues, in time
$O((2^k\log n)(\log U)/k)$.
\end{proof}

We summarize the results of this section:

\begin{theorem}\label{thm:dyn-int}
For any $k$ there exists a data structure for maintaining dynamic
prioritized intervals in the range $[0,U-1]$, and finding the maximum
priority interval containing a query point in any persistent version of
the data structure, in time
$O((\log U)/k)$ per query, time
$O((2^k\log n)(\log U)/k)$ per update,
and space $O(2^k(\log U)/k)$ per update.
\end{theorem}

The $\log n$ factor in the update time can be reduced by building a
segment tree of subintervals within each block, and maintaining a
priority queue of the dynamic intervals corresponding to each canonical
interval of the segment tree; we omit the details, since this factor does
not form an important part of our overall running time and can (as
detailed below) be avoided entirely for hierarchical rectangles.

\subsection{Static to Dynamic Transformation}

We use the dynamic data structure of the previous subsection to solve
our static rectangle querying problem, as follows.

\begin{lemma}\label{lem:int-pred}
Suppose we are given a set $S$ of $n$ integers in the range $[0,U-1]$.
Then for any $x$ we can build a data structure which finds the largest
predecessor in $S$ of a given query integer, in space $O(nx\log_x U)$ and
query time
$O(\log_x U)$.
\end{lemma}

\begin{proof}
Form a set of intervals $[i,U-1]$ with priority $i$ for $i\in S$.
The maximum priority interval containing $q$ has as its left endpoint
the predecessor of $q$.  Thus, we can use a static version of the data
structure described in Theorem~\ref{thm:dyn-int} (with $k=\log x$) to
solve this problem.
\end{proof}

\eject

Beame and Fich~\cite{BeaFic-STOC-99} provide matching $\Theta(\log\log n
/ \log\log\log n)$ upper and lower bounds for integer predecessor
queries in polynomial space, and survey several previous results on the
problem.  Because of the reduction above, their lower bounds apply as
well the the maximum priority interval and rectangle problems.  Our
results escape this lower bound by having a space bound that
depends on $U$ and not just on $n$.

\begin{theorem}
Given a set of $n$ axis-aligned prioritized rectangles with coordinates in
the range
$[0,U-1]$, and a parameter $x$, we can build a data structure of size
$O(nx\log_x U)$ which
can find the maximum priority rectangle containing a query point
in time $O(\log_x U)$.
\end{theorem}

\begin{proof}
We consider a left-right sweep of the rectangles by a vertical line; for
each position of the sweep line we maintain a dynamic set of intervals
formed by the intersections of the rectangles with the sweep line.
This intersection changes only when the sweep line crosses the left or
right boundary of a rectangle; at the left boundary we insert the
$y$-projection of the rectangle and at the right boundary we delete it.
With each rectangle boundary we store a pointer to the version of the
data structure formed when crossing that boundary.

A query can be handled
by using the integer predecessor data structure of
Lemma~\ref{lem:int-pred} to find the
$x$-co\-ord\-in\-ate of the nearest rectangle boundary to the right of the
query point, and then performing a query in the corresponding version of
the interval data structure.
\end{proof}

In particular when $U=n^{O(1)}$ we achieve
query time $O(\log\log n)$ in space $O(n^{1+o(1)})$,
or query time $O(1)$ in space $O(n^{1+\epsilon})$,
while previous solutions used
space $\Theta(n^{1+\epsilon})$ to achieve query time $O(\log\log n)$
\cite{FerMutdeB-STOC-99}.

It is not difficult
to modify our data structure to handle other decomposable queries, such as
listing all rectangles containing the given query point, in similar time
and space bounds.

For hierarchical rectangles, we can simplify the dynamic interval data
structure by using insertion and undo operations instead of more general
insertions and deletions, and by omitting the ${\rm pq}[i]$ pointers and
the priority queues they point to.  An insertion can be handled by
comparing the priority of the newly inserted interval to the values
${\rm opt}[i]$ for the blocks containing the interval's endpoints.
An undo can be handled simply by restoring the pointer to the top-level
block to its previous version.

\section{Conflict Detection}
\label{sec:conflict}

We say that a set of rules, represented by a set of rectangles with
priorities, has a {\em conflict} if there exists a query point $q$ such
that there is not a unique maximum-priority rectangle containing $q$. 
Note that this is a stronger condition than the existence of an
intersecting pair of equal-priority rectangles, since a higher-priority
rectangle could cover the intersection and avoid a conflict.
As defined in the introduction, the filter conflict detection problem
further restricts conflicts to rules with conflicting actions; the
algorithms described here can be extended to cases where the actions
can be partitioned into a small number of conflict types
but we omit the details.

We would like to know whether a given set of prioritized rectangles has
a conflict.  A naive method would test each pair of equal-priority
rectangles to determine whether they conflict, but this would not be
efficient due to the difficulty of testing whether their intersection is
covered by the union of higher priority rectangles.  Less naively, the
problem can be solved in near-quadratic time by querying each point
determined by the horizontal boundary of one rectangle and the vertical
boundary of another, or by constructing the arrangement of all the
rectangles and using a priority queue to find the maximum priority
rectangle(s) within each arrangement cell.  We seek an even more
efficient (subquadratic) solution.

\subsection{Priority Queues, Lower Envelopes, and\\ MST Verification}

Consider the following three problems:

\begin{figure*}
\begin{center}
\includegraphics{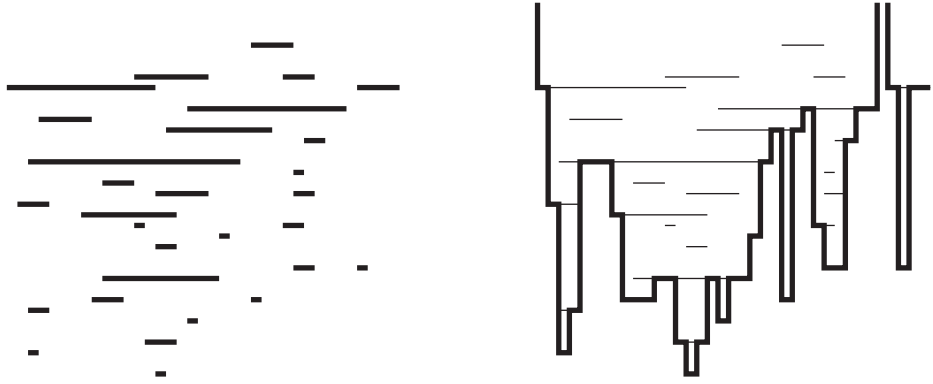}
\end{center}
\caption{A set of horizontal line segments (left) and its lower envelope
(right).}
\label{fig:lowerenv}
\end{figure*}

\begin{itemize}
\item Given is an offline sequence of $O(n)$ integer priority queue
operations: insert or delete a value in the set $\{0,1,\ldots n-1\}$
and query the minimum value.
How quickly can one answer all the queries?

\item Given is a set of horizontal line segments
(Figure~\ref{fig:lowerenv}, left), each endpoint of which has coordinates
in the set $\{0,1,\ldots n-1\}$. How quickly can one construct the lower
envelope of the line segments? That is, if we think of each line segment
as representing the graph of a (constant) function defined over a portion
of the $x$-axis, what is the (piecewise constant) minimum of these
functions (Figure~\ref{fig:lowerenv}, right)?

\item Given is a graph, in which the minimum spanning tree is a given
path, and in which all edges have weights in the set
$\{0,1,\ldots n-1\}$.  How quickly can one determine, for each edge in
the path, which edge would replace it in the MST if the path edge were
deleted?
\end{itemize}

It is not difficult to see that in fact these problems are equivalent to
each other: the insertion times, deletion times, and priorities in the
offline priority queue correspond respectively to the $x$-coordinates of
the left endpoints, $x$-coordinate of the right endpoints, and
$y$-coordinates of the horizontal line segments, which correspond
respectively to the first vertex (according to the path order), second
vertex, and weight of the non-MST edges in the graph.

Aho, Hopcroft, and Ullman~\cite[pp. 139--141]{AhoHopUll-74} describe an
algorithm for a similar offline priority queue problem, however their
problem involves delete-minimum operations rather than deletions of
particular values. Although the best replacement edge for each non-MST
edge can be found in linear time~\cite{Kin-Algo-97}, the fastest known
algorithm for finding the best replacement for each MST edge (without the
integer restriction) remains Tarjan's slightly superlinear
one~\cite{Tar-JACM-79}.

\begin{lemma}\label{lem:horiz-envelope}
The three problems described above can be solved in linear time.
\end{lemma}

\begin{proof}
We consider the minimum spanning tree verification formulation of the
problem, and consider the non-tree edges in sorted order by weight.
Our algorithm finds the replacements for each path edge in a certain
order; when a path edge's replacement is found we reduce the size of the
graph by contracting that edge.  This contraction clearly does not
change the replacement for the remaining edges.  We use a union-find
data structure to keep track of the relation between the original graph
vertices and the vertices of the contracted graph.  Since the
contractions will be performed along the edges of a fixed tree (namely,
the given path), we can use the linear-time union-find data structure of
Gabow and Tarjan~\cite{GabTar-JCSS-85} or its recent simplification by
Alstrup et al.~\cite{AlsSecSpo-IPL-97}.

Our algorithm, then, simply performs the following steps for each edge
$(u,v)$, in sorted order by edge weight:
for each uncontracted edge $(x,y)$ remaining in the path between $u$ and
$v$, set that edge's replacement to $(u,v)$, contract the edge,
and unite $x$ and $y$ in the union-find data structure.

The time per edge $(u,v)$ is a constant, plus a term proportional to the
number of path edges contracted as a result of processing edge $(u,v)$. 
Since each edge can only be contracted once, the total time is linear.
\end{proof}

The technique readily extends to finding best replacement edges for graphs
where the MST is not a path.

For our application to conflict detection, we also need to know whether
there were any ambiguities in the above process; that is, whether any of
the offline min operations can return more than one equal minimum value.
This is essentially the same as our original conflict detection problem
in one dimension rather than two. One way to solve this is to apply
the above algorithm twice, once with an arbitrary tie-breaking order
imposed on equal weight edges, and once again with the reverse order
imposed, and test whether the two applications of the algorithm produce
the same assignment of replacement edges.

\subsection{Stripes}

We first describe an efficient algorithm for conflict detection in the
special case that each rectangle is a {\em stripe}; that is, either its
vertical extent or its horizontal extent is the entire space $[0,U-1]$.
We do not expect such a restricted case to occur in our application, but
it forms an important subroutine for our more general algorithm.

We classify stripes into three types:
\begin{itemize}
\item A {\em horizontal} stripe has $x$-extent $[0,U-1]$
and $y$-extent a proper subset of $[0,U-1]$.
\item A {\em vertical} stripe has $x$-extent a proper subset of
$[0,U-1]$ and $y$-extent $[0,U-1]$.
\item A {\em universal} stripe has both $x$-extent and $y$-extent
equal to the entire space $[0,U-1]$.
\end{itemize}

\begin{lemma}\label{lem:stripes}
Let a collection of prioritized stripes be given, together with sorted
orderings of all stripes according to their priorities, the horizontal
boundaries of horizontal stripes according to their $y$-coordinates, and
the vertical boundaries of vertical stripes according to their
$x$-coordinates. Then we can detect a conflict in this set of stripes in
linear time.
\end{lemma}

\begin{proof}
We first partition the space $[0,U-1]^2$ into horizontal stripes,
according to the maximum-priority horizontal input stripe covering each
point in the space; essentially this is just the lower envelope
computation of Lemma~\ref{lem:horiz-envelope}.
Let $m_h$ denote the minimum priority occurring in this partition.
Similarly, we partition the space into vertical stripes
according to the maximum-priority vertical input stripe covering each
point, and let $m_v$ denote the minimum priority occurring in this
partition. Finally, we let $m_u$ denote the maximum priority of any
universal stripe.  (We set $m_h$, $m_v$, or $m_u$ to $-\infty$ if
the corresponding set of stripes is empty.)

We then use this information to search for conflicts, as
follows, depending on the types of the two conflicting stripes:
\begin{itemize}
\item To find a conflict between two horizontal stripes, if one exists,
test whether there exists an ambiguity in the construction of the
horizontal partition, as discussed below Lemma~\ref{lem:horiz-envelope}. 
If there is such an ambiguity, let $p_h$ denote the maximum priority of
any ambiguity.  Then a conflict exists if and only if
$p_h\ge\max\{m_v,m_u\}$.
Similarly we can find a conflict between two vertical stripes
by letting $p_v$ denote the maximum priority of an ambiguity in the
vertical partition, and testing whether $p_v\ge\max\{m_h,m_u\}$.
\item A conflict between two universal stripes exists if and only if
some two or more universal stripes have priority $m_u$,
and if $m_u\ge\max\{m_h,m_v\}$.
\item A conflict between a universal and a horizontal stripe
exists if and only if $m_u$ is also the priority of one of the stripes in
the horizontal partition, and $m_u\ge m_v$.  Similarly a conflict
between a universal and a vertical stripe
exists if and only if $m_u$ is also the priority of one of the stripes
in the vertical partition, and $m_u\ge m_h$.
\item A conflict between a horizontal stripe and a vertical stripe
exists if and only if there is a priority $p\ge m_u$ that appears both
in the horizontal and the vertical partition.
\end{itemize}

Thus, the problem has been reduced to a constant number of comparisons,
together with two more complex operations: determining whether $m_u$
appears in either of two sets of priorities, and determining the
intersection of those two sets.  Since we know the sorted order of the
priorities, we can represent them by values in the range $[0,n-1]$
and use a simple bitmap to perform these membership and intersection
tests in linear total time.
\end{proof}

\subsection{kD-tree}

\begin{figure*}[p]
\begin{center}
\centerline{\hbox{\includegraphics[width=3in]{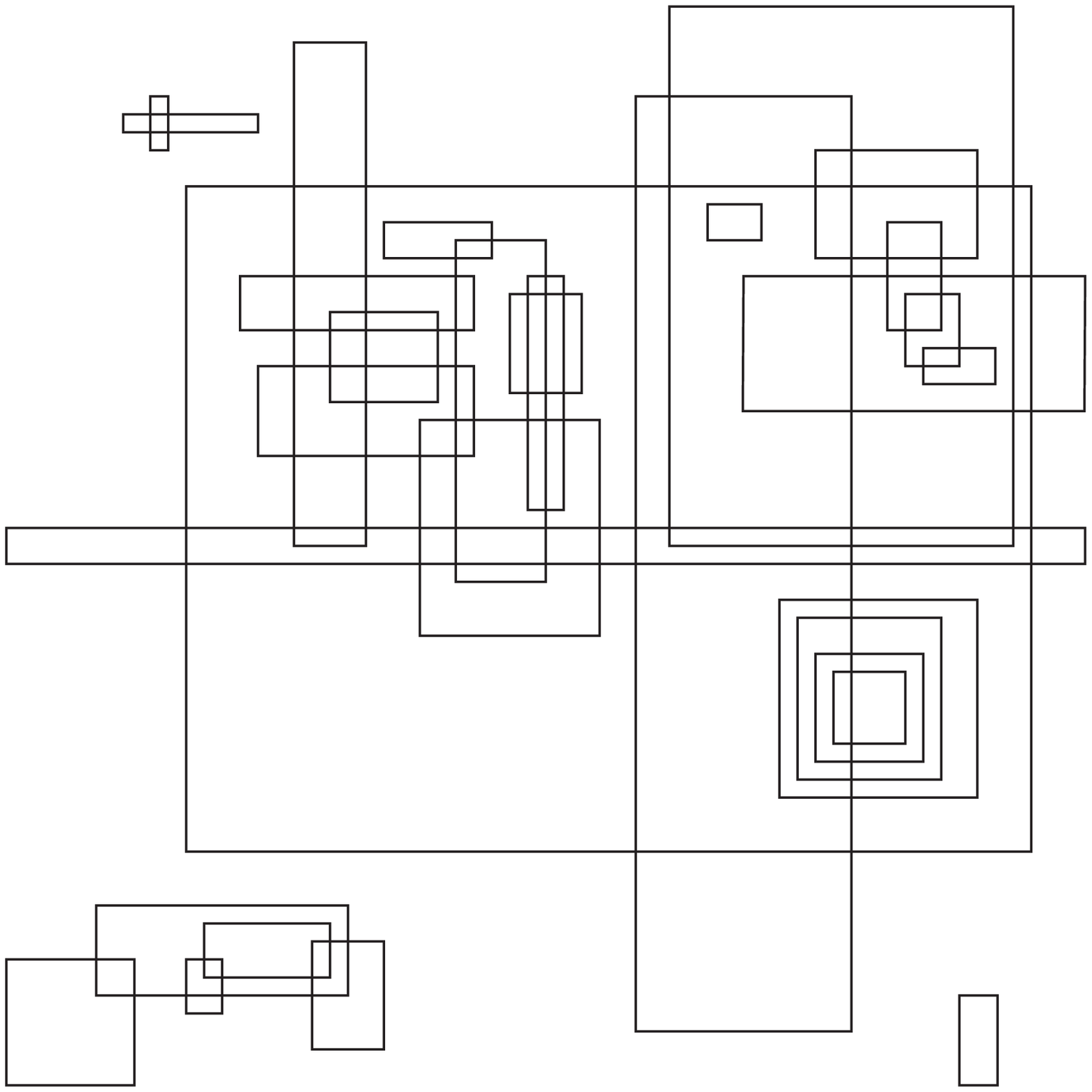}
\includegraphics[width=3in]{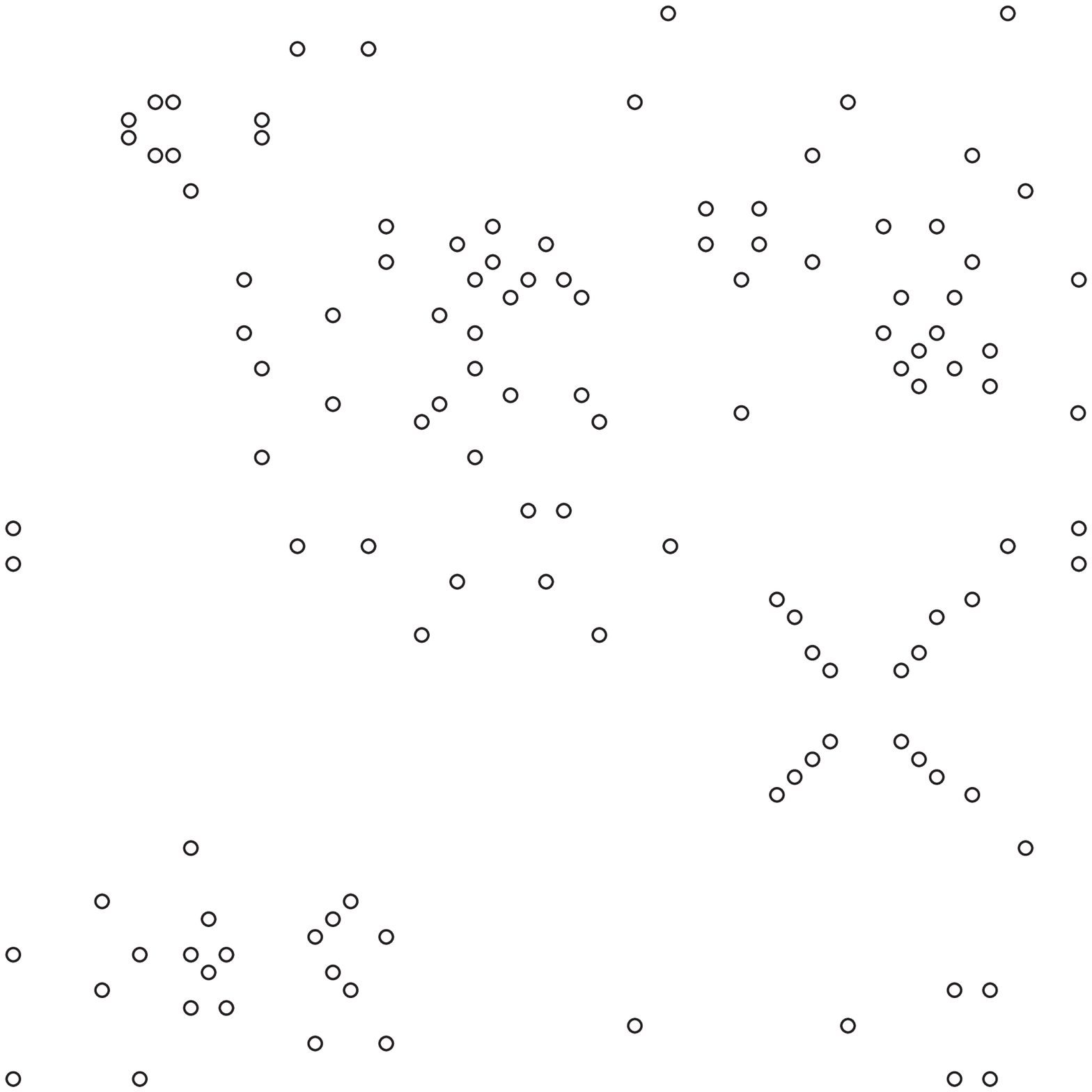}}}

\centerline{\hbox{\includegraphics[width=3in]{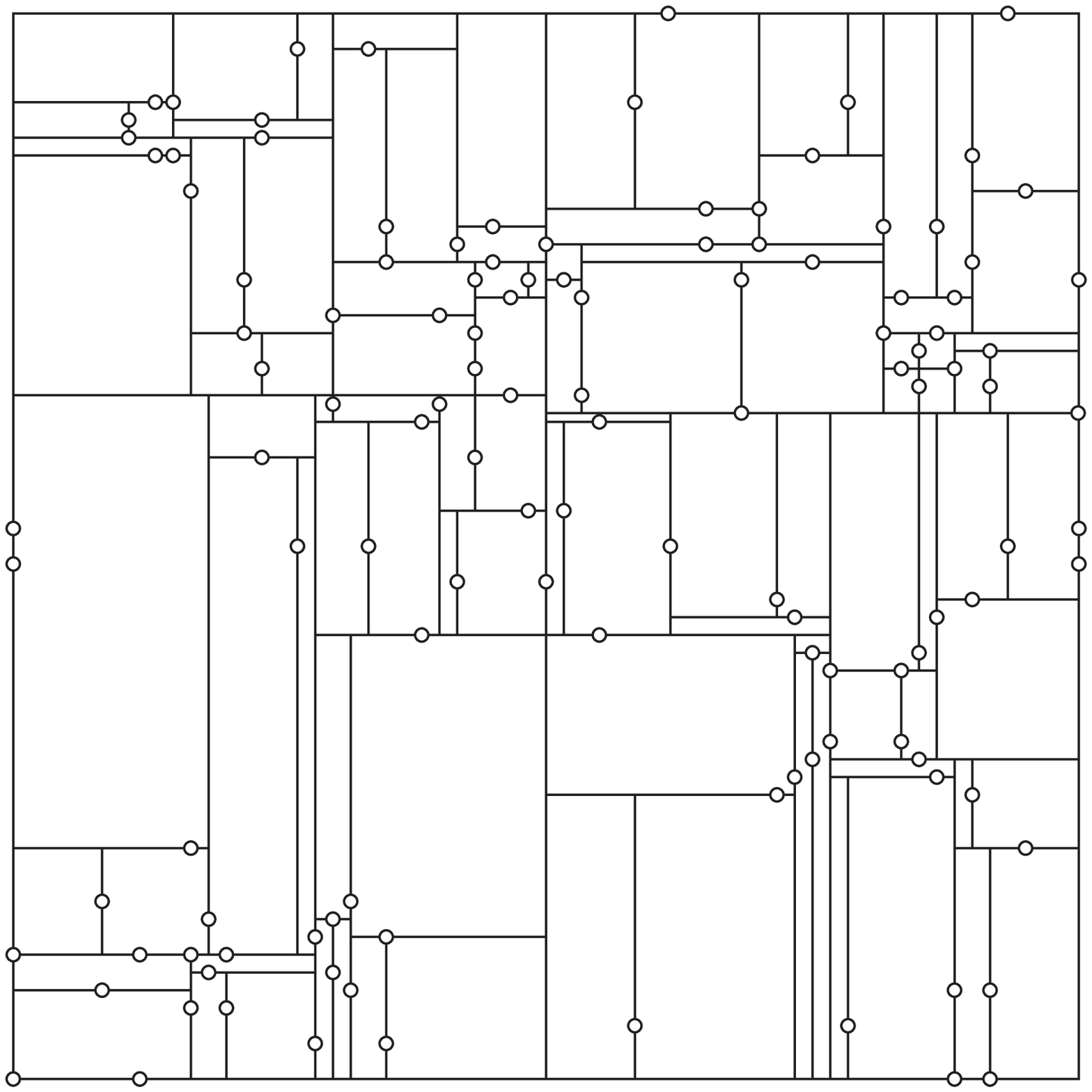}
\includegraphics[width=3in]{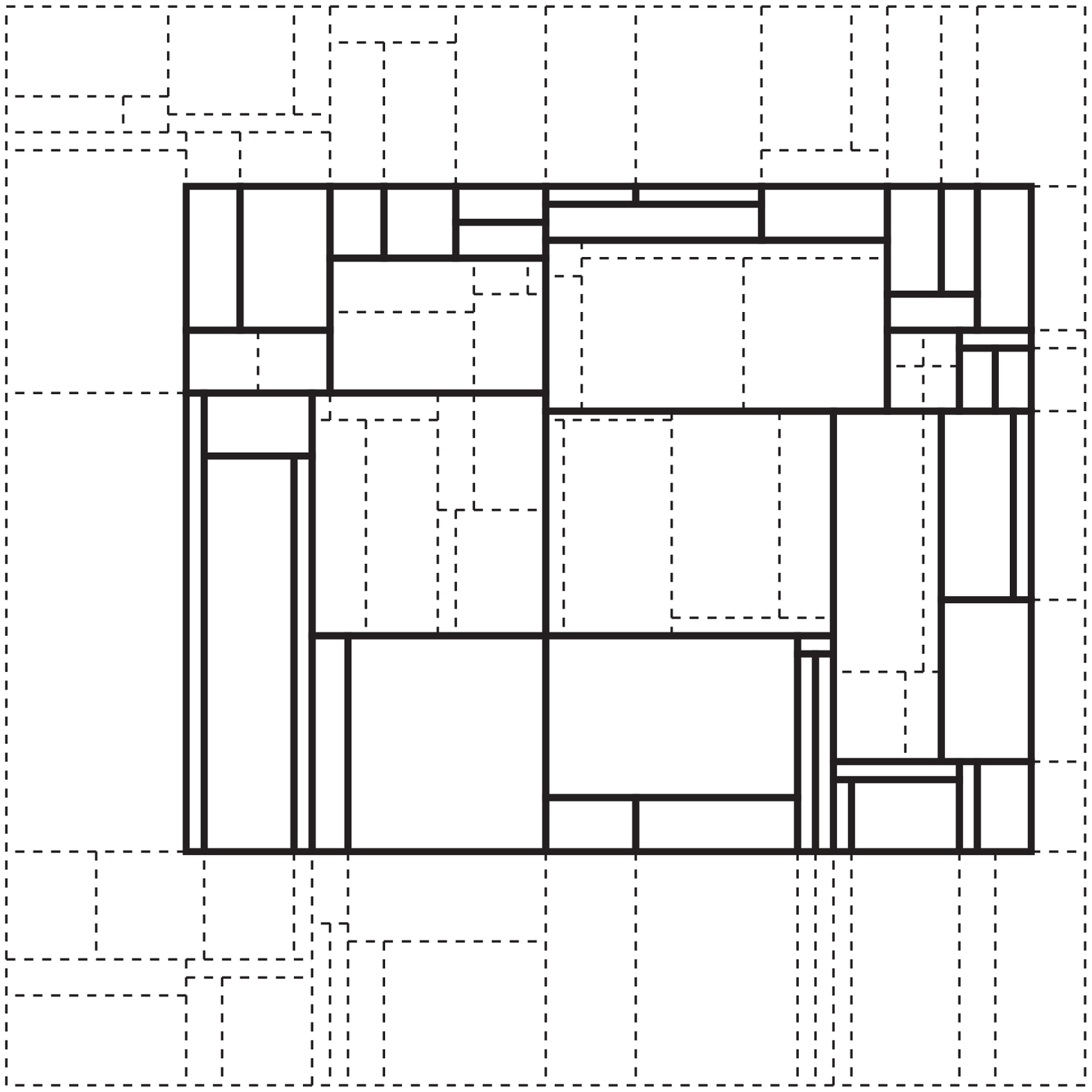}}}
\end{center}
\caption{$kD$-tree for a set of rectangles.  Upper left: the rectangles.
Upper right: their vertices.  Lower left: $kD$-tree for the vertices.
Lower right: maximal $kD$-tree cells partition an input rectangle.}
\label{fig:kdtree}
\end{figure*}

A {\em $kD$-tree} \cite{Ben-CACM-75} of a set of points is a hierarchical
partition into rectangular {\em cells}, formed as follows:
\begin{itemize}
\item The root of the hierarchy is a bounding box for the point set.
\item If a cell at an even level of the hierarchy contains
one or more points in its interior, then it is split into two smaller
cells by a vertical line through the point with the median
$x$-coordinate.
\item If a cell at an odd level of the hierarchy contains
one or more points in its interior, then it is split into two smaller
cells by a horizontal line through the point with the median
$y$-coordinate.
\end{itemize}

If a cell contains an even number of points, either of the two median
points can be used to determin its split line. The leaf cells of the
$kD$-tree form a partition of the bounding box into
$O(n)$ empty rectangles.  Since each split divides its point set in
half, the number of levels of the hierarchy is at most $\log_2 n$
(it can be smaller if several points are contained in a single split
line).

\begin{lemma}\label{lem:kdt-line}
Any vertical or horizontal line cuts $O(\sqrt n)$ cells at all levels of
the $kD$-tree for a set of $O(n)$ points.
\end{lemma}

\begin{proof}
If the line is horizontal (vertical), the number of cells cut by the line
at most doubles at every even (odd) level of the $kD$-tree construction,
and remains unchanged at every odd (even) level.  The result follows
from the $\log_2 n$ bound on the number of levels in the tree.
\end{proof}

In our conflict detection algorithm, we will use $kD$-trees
defined on the set of corners of the input rectangles
(Figure~\ref{fig:kdtree}).  We say that an input rectangle {\em covers}
a $kD$-tree cell if the cell is completely contained in the rectangle.
We define a {\em maximal covered cell} for a given rectangle to be a
cell that is covered by the rectangle, but for which the cell's parent
is not covered.  We say that a rectangle {\em crosses} a cell
if it has a nonempty intersection with the interior of the cell but does
not cover it.

\begin{lemma}\label{lem:kdt-rect}
Any rectangle has $O(\sqrt n)$ crossed cells and $O(\sqrt n)$ maximal
covered cells at all levels in the $kD$-tree.
\end{lemma}

\begin{proof}
The bound on the number of crossed cells follows immediately from
Lemma~\ref{lem:kdt-line}. The parent of a maximal covered cell must be
crossed, and each crossed cell can have at most one maximal covered
child, so the number of maximal covered cells is also $O(\sqrt n)$.
\end{proof}

\subsection{The Conflict Detection Algorithm}

Clearly, if a set of rectangles has a conflict, then this conflict must
occur within at least one of the leaf cells of a $kD$-tree.
Further, since the leaf cells contain no rectangle corners, each
rectangle acts like a stripe within any such cell: it extends either the
full width or the full height of the cell.  Thus, we can perform
conflict detection by building a $kD$-tree and applying our stripe
conflict detection algorithm to each cell.

\begin{theorem}
Given a set of $n$ prioritized rectangles, we can determine whether the
set has a conflict in time $O(n^{3/2})$ and space $O(n)$.
\end{theorem}

\begin{proof}
We build a $kD$-tree of the rectangle vertices (this can be done in time
$O(n\log n)$ and perform a depth first traversal of the tree.
As we traverse the tree, we maintain at each cell of the traversal the
following information:
\begin{itemize}
\item The maximum priority of a rectangle covering the cell,
and one or (if they exist) two rectangles having that maximum priority.
\item A list of the rectangles crossing the cell, sorted by priority.
\item A sorted list of the horizontal boundaries of the rectangles
that cross the cell.
\item A sorted list of the vertical boundaries of the rectangles
that cross the cell.
\end{itemize}

When the traversal reaches a cell $C$, we can determine which of
rectangles cross or cover the children of $C$, and extract the sorted
sublists for its two children, in time linear in the number of
rectangles crossing $C$.  We also find the set of rectangles that cross
$C$ but maximally cover one of its children, scan this set for the
maximum priority, and use this information (together with the maximum
priority of a rectangle covering $C$) to determine the maximum priority
of a rectangle covering each child.

When the traversal reaches a leaf
cell, we apply the algorithm of Lemma~\ref{lem:stripes} to test whether
the cell contains a conflict.

While one child of a cell $C$ is being
processed recursively, we store with
$C$ only the portions of the sorted lists that have not been passed to
that child, so that each rectangle or rectangle edge is stored in one of
the lists only at a single level of the tree, keeping the total space
linear. All operations performed when traversing a cell take time linear
in the number of rectangles crossing or maximally covering the cell,
so by Lemma~\ref{lem:kdt-rect} the total time is $O(n\sqrt n)$.
\end{proof}

\section{Concluding Remarks}
\label{sec:conc}

We have considered the two fundamental packet filter management
problems in IP networks, namely, packet classification and filter 
conflict detection, for the two dimensional case of immediate interest. 
For the packet classification problem,
we present a simple algorithm that takes $O(\log\log n)$ 
time to classify packets matching the best known bounds for the 
one dimensional case, and improving upon the
space needed by currently known solutions. 
For the filter conflict detection problem, our solution is the 
first sub-quadratic time algorithm.
 
Our packet classification algorithm may well turn out to be 
better than existing ones in practice, too. 
We fully intend to test that possibility. However, 
the task is not one of merely implementing our algorithm and 
comparing against the known ones. Since the study of 
packet classification is quite mature in the networking communities,
we need to do a careful job adapting our solution (where to
make best use of large memory cache line, how to combine hardware
and software solutions, how to exploit the properties of rule sets
to isolate small, hard subproblems where our solution
will be useful, etc). Engineering such tradeoffs is best 
explored in a separate paper. 

Dynamic versions of the packet filter management problem are open, as
are extensions of our results to higher dimensional query problems.

\bibliographystyle{abbrv}
\bibliography{rectangles}

\end{document}